# Spatial Modelling Techniques in Microsoft Excel

Stephen Allen
ACBA (UK) LTD
steve@acba.co.uk

**ABSTRACT**

*We begin by considering the expectations of the creators of VisiCalc, the first spreadsheet. The emphasis is on the nature of the spreadsheet grid. The grid is taken as a presentational method for showing a solution to a Sudoku puzzle. We consider methods or approaches for the solution.*

*We look at the relationship between this model and academic papers on the methods for describing and categorising end-user models generally. We consider whether the type of model described here should be categorised separately. The complexity of the model is reviewed in the context of commendations to minimise overly sophisticated presentational constructs and formulae.*

## 1. BACKGROUND

### Nature and Role of a Spreadsheet

The philosophical question of the core nature of a spreadsheet is not prominent in the literature. Its practical uses are so obvious that we have just got on and used it – for good or ill. However, Bricklin and Frankston, the spreadsheet creators [Power, 2004], did not consider it primarily as an accountancy program: as was the case with the first spreadsheet-type concepts built for mainframe computers. In particular, Frankston stated in an email to Power:

"The grid provided the simplifying structure that made it a spreadsheet as opposed to a more general surface."

The importance of the grid, as opposed to a table or similar database like structure, is huge. It allows spatial freedom with the placement of the grid and conceptual freedom over the nature and contents of the grid (e.g. numbers, text, formulae, images, colours etc.)

The grid structure is often obscured by other elements. There is a great deal of redundancy between the Microsoft business programs [Cook, 2010]. Primarily, Word is for word processing, Excel is the spreadsheet, Access is the database etc. But Excel has database features, Word has spreadsheet features, etc.

The emphasis on database functions within spreadsheets has obscured the functional elements of the grid style paradigm. This paper contributes to redressing the balance, by:

- Selecting a model that employs a grid structure to determine the relationship between individual identities
- Using numeric identities rather than number values, thus forcing a different programmatic approach to the management of the model

We selected the Sudoku puzzle as our experimental model. This model has two additional benefits:

- It involves a relatively complex grid structure where columns and rows and sub-grids are superimposed on top of one another
- We could create a modelling approach that allowed the user to select his own personal pathway for solving the puzzle

### 1.1. Sudoku as a Model

The objective of the puzzle is to fill a 9×9 grid with digits so that each column, each row, and each of the nine 3×3 sub-grids that compose the grid (also called "boxes", "blocks", "regions", or "sub-squares") contains all of the digits from 1 to 9. The puzzle setter provides a partially completed grid, which typically has a unique solution.

A Sudoku grid, like a spreadsheet, is a two dimensional object. Accordingly it is easy to present the puzzle to the user in this medium and identify each of the rows and columns (Figure 1).

**Figure 1 – A Sudoku puzzle from The Guardian, 18 February 2012**

**The Original Grid**

| | | | | | | | | |
|---|---|---|---|---|---|---|---|---|
| . | . | . | . | . | . | . | . | . |
| 2 | 3 | . | 6 | . | . | . | 1 | . |
| 4 | 5 | . | . | . | 3 | 6 | . | . |
| . | . | . | 2 | . | . | . | 9 | 8 |
| 6 | 2 | . | . | . | . | . | 7 | 4 |
| 3 | 4 | . | . | . | 7 | . | . | . |
| . | . | 2 | 5 | . | . | . | 4 | 9 |
| . | 7 | . | . | . | 6 | . | 5 | 2 |
| . | . | . | . | . | . | . | . | . |

In Sudoku the numbers simply represent identities. The true purpose of the puzzle was to follow a series of very specific and relatively simple logic requirements associated with the individual identity in relation to the spatial elements of the grid.

The Sudoku puzzle employs overlapping ranges. It stipulates logical relationships based entirely on the overlaps such that it can be regarded as a three dimensional model in a two dimensional space. The capacity of spreadsheets to cope with overlapping ranges [Bricklin, 1999] was a major advantage and is fully utilised in this context.

### 1.2. The Structure of the Paper

Overall, this approach to solving Sudoku puzzles is fairly extended and complex. For a more comprehensive history and detailed description see [Allen, 2012a]. The spreadsheet styles and constructs necessary for the development of this conceptual overview will be divided over two papers. This first will deal with the methods and approaches adopted for the analysis of individual cells within the grid. A second paper will consider methods and approaches for changing the focus of cells under examination. These measures are designed to demonstrate the specific pathway through which all the Sudoku cells are evaluated.

This paper will consider the styles of analytical approach. The detail (Section 2) considers the essential decision making part of the design structures. We examine the relationship between the structure, the potential areas at risk of error and the control mechanisms adopted to manage those risks.

We take the opportunity here to consider the models in relation to previous academic papers that concentrate on the structural description of spreadsheets (Section 3) and draw tentative conclusions. The Sudoku model looks simple, but the overlapping grids hide an inherent complexity. We discuss this complexity in the light of Bregar's [2004] paper.

## 2. THE MAIN ANALYTICAL MODELS

### 2.1. Introduction

We need two primary templates because there are two major methodologies for solving a Sudoku puzzle. One seeks to eliminate potential solutions for a particular cell until the player is left with only one. The other considers the relationships for individual identities within a group of three related columns or rows and their associated sub-grids. We also introduce a third template, but this depends on the output from the template that seeks to eliminate potential solutions. It is of lesser importance than the primary templates.

The basis for both the primary modelling templates was that the user would make a decision about the value of an individual Sudoku cell and demonstrate/record the rationale for that decision.

In each case we describe the primary decision and the major controls that the model employs to stop wholly irrational decisions that would wreck the model. We sought to draw a balance between decisions that were incorrect but were acceptable within the context of the model and those that would undermine the integrity of the model as a whole.

### 2.2. Analysis by Exclusion

The theory of this approach is that the user selects a cell within the puzzle that has yet to be solved. The role of the user is to eliminate all identities bar one by reference to their pre-existence in any one of the three dimensions directly related to that cell. In this context, the system automatically evaluates which numeric identities have already been employed within the row dimension, leaving the user to demonstrate the existence of identities in the other two dimensions.

In Figure 2, the cells shaded in background green represent the user input – the columns headed "Col" and "Grid". The content of each cell, within this range, is restricted to 9 potential references which refer to an individual cell position within the Sudoku grid. In effect, the user is expressing the view that the cell under consideration cannot contain this identity because it already exists elsewhere within a related dimension. The value that a user employs is designed to show exactly where the identity is located within that related dimension.

We emphasise that this is a user decision, not an automatic one. There remain many possible reasons why the user decision can be incorrect. The point of the model is to record the rationale, so that a reviewer can validate it. This is a crucial element in the control methodology, which automates certain processes (e.g. posting a value to a Sudoku cell). It is the user's logic that we record. The system evaluates the result on the basis of that logic.

**Figure 2– Both automated and user input to the analysis of a Sudoku cell**

| Column 2 | | Status | | Open | | Col3 |
|---|---|---|---|---|---|---|
| | | | | | | |
| | Avail | Col | Grid | AvailNow | | |
| 1 A | y | 4_24 | | n | | |
| 2 B | y | 4_25 | | n | | |
| 3 C | | | | n | | |
| 4 D | | | | n | | |
| 5 E | y | 1_23 | | n | | |
| 6 F | | | | n | | |
| 7 G | y | 7_28 | | n | | |
| 8 H | y | | 7_17 | n | | |
| 9 I | y | | | y | | 9 |
| | Count "y" | | | 1 | | 9 |
| | | | | | | |
| | Result | | | Solved | | 9 |

Commentary

Once the player has posted his analysis of the review of the column and sub-grid dimensions the evaluation of the results is automated. The process is both stepwise and visible (Figure 2). "AvailNow" is (effectively) a Boolean formula and shows whether the number identity for the cell is still available. The cell, two positions to the right, specifies the numeric identity. The group of 4 cells at the bottom of the analysis grid summarise the user's evaluation inputs.

The user input to the analytical process is strictly limited, but one of the positive impacts of the process is to expose clearly both the procedures and logic of generating a conclusion.

### 2.3. Location Analysis for a Specific Value

Since the three dimensions (row, column and sub-grid) are superimposed on top of one another, when a user selects a row/column/sub-group triplet, he should find a single identity repeated three times such that it is present once in each of the three dimensions. This characteristic can be employed for analytical purposes, where two of the locations of an individual identity are already known. In such a case the potential location for the third member of the group is limited to three identifiable cells.

**Figure 3 – Location Analysis for identity '2' – detail from the middle three rows and associated sub-grids**

| | | | | | | | | |
|---|---|---|---|---|---|---|---|---|
| . | 1 | . | 2 | . | . | . | 9 | 8 |
| 6 | 2 | . | . | . | . | . | 7 | 4 |
| 3 | 4 | . | . | . | 7 | . | 6 | 1,6 |

The overall analysis (whether automated or through user input) seeks to eliminate two of the three potential locations, either because it is already solved or because the identity already exists in an associated dimension of the individual cell. The user input is limited to identifying those cells that can be excluded as a potential solution because the identifier is present in a related dimension (Figure 4)

**Figure 4 – User input for location analysis**

| | H | I | J | K | L | M | N | O | P | Q | R | S | T | U | V |
|---|---|---|---|---|---|---|---|---|---|---|---|---|---|---|---|
| 38 | | | | | | | | | | | | | | | |
| 39 | | Possible Solutions | | | | Col Analysis | | | | Excluded? | | | | Result | |
| 40 | | 6_76 | | . | 6 | 7 | | | | | n | | Calc | 2 | |
| 41 | | 6_86 | | 6 | 6 | 8 | Solv | | | | y | | | | |
| 42 | | 6_96 | | 1,6 | 6 | 9 | | 6_96 | | | y | | | | |
| 43 | | | | | | | | | | | | | | | |
| 44 | | | | | | | Count Exclusions | | | | 2 | | | | |

Cell R44 (Figure 4) counts the number of locations excluded and determines the nature/ presentation of the final result. The most useful result solves the value of a single cell. Once again the system is designed to record the user's logic and interprets the result automatically.

## 2.4. Mutual Exclusion Analysis

This analysis depends on the '1 of 2' output from "Analysis by Exclusion" process (Section 2.2). It relies on the concept that if, within a specified dimension, two Sudoku cells have identical values for their '1 of 2' output then BOTH the identities must be associated with those two cells exclusively. Accordingly, if a third cell claims one of the original two values as an alternative, we can assert with certainty that the value is not available because of the mutual exclusivity claimed by the original two cells. This is better illustrated by a diagram (Figure 5).

**Figure 5 – Output from Mutual Exclusion Analysis**

| | | | | | | | | | | | | | | | | | | |
|---|---|---|---|---|---|---|---|---|---|---|---|---|---|---|---|---|---|---|
| Analysing Alternative Solutions for Columns | | | | | | | | | | | | | | | | | | |
| Examining column | | | | | | | | 8 | B006 | | | | | | Original Source | | | F051 |
| 1 | 2 | 3 | 4 | 5 | 6 | 7 | 8 | 9 | | | | | | | | | | B/Fo |
| Currently Available Solutions | | | | | | | | | Grid Current Status | | | | | | | | | F066 |
| 1 | 2 | | 4 | 5 | 6 | 7 | 8 | 9 | . | 8,9 | . | . | . | . | . | 2,8 | 3 | Res |
| | | | 4 | | | 7 | 8 | 9 | 2 | 3 | . | 6 | . | . | . | 1 | 5 | |
| 1 | 2 | | | | | | 8 | 9 | 4 | 5 | . | . | . | 3 | 6 | 2,8 | 7 | |
| | | 3 | 4 | 5 | 6 | 7 | | | . | 1 | . | 2 | . | . | . | 9 | 8 | |
| 1 | | 3 | | 5 | | | 8 | 9 | 6 | 2 | . | . | . | . | . | 7 | 4 | |
| 1 | 2 | | | 5 | 6 | | 8 | 9 | 3 | 4 | . | . | . | 7 | . | 2,6 | 1,6 | |
| 1 | | 3 | | | | 7 | 8 | | 1,8 | 6 | 2 | 5 | 3,7 | 1,8 | 3,7 | 4 | 9 | |
| 1 | | 3 | 4 | | | | 8 | 9 | . | 7 | . | . | . | 6 | . | 5 | 2 | |
| 1 | 2 | 3 | 4 | 5 | 6 | 7 | 8 | 9 | . | 8,9 | . | . | . | . | . | . | . | |
| | | If total is odd number - systems error | | | | | | | | 2 | | | | | | | | C/Fo |
| | | | | | | | | | | | | | | | | B006 | | |
| Revised Available Solutions | | | | | | | | | Grid Revised Status | | | | | | | | | F068 |
| 1 | 2 | | 4 | 5 | 6 | 7 | 8 | 9 | . | 8,9 | . | . | . | . | . | 2,8 | 3 | |
| | | | 4 | | | 7 | 8 | 9 | 2 | 3 | . | 6 | . | . | . | 1 | 5 | |
| 1 | 2 | | | | | | 8 | 9 | 4 | 5 | . | . | . | 3 | 6 | 2,8 | 7 | |
| | | 3 | 4 | 5 | 6 | 7 | | | . | 1 | . | 2 | . | . | . | 9 | 8 | |
| 1 | | 3 | | 5 | | | 8 | 9 | 6 | 2 | . | . | . | . | . | 7 | 4 | |
| 1 | 2 | | | 5 | | | 8 | 9 | 3 | 4 | . | . | . | 7 | . | 6 | 1,6 | |
| 1 | | 3 | | | | 7 | 8 | | 1,8 | 6 | 2 | 5 | 3,7 | 1,8 | 3,7 | 4 | 9 | |
| 1 | | 3 | 4 | | | | 8 | 9 | . | 7 | . | . | . | 6 | . | 5 | 2 | |
| 1 | 2 | 3 | 4 | 5 | 6 | 7 | 8 | 9 | . | 8,9 | . | . | . | . | . | . | . | |

Figure 5 shows the before and after effects of the Mutual Exclusion Analysis. The only user input is to identify the dimension for examination – in this case Column 8.

The worksheet automatically notes that the first and third cells in this column contain the identical values '2, 8'. These two integer identities are therefore deemed unavailable to the other cells within the Column 8 dimension. Nevertheless, the sixth cell in this column claims potential alternatives of '2, 6'. Since the identity '2' is claimed on a mutually exclusive basis, by the first two cells the system knows that the result of that cell must be the identity '6'. It is illustrated accordingly in bold green.

## 3. SPREADSHEET STRUCTURES AND RISKS

### 3.1. Methods of Building a Spreadsheet

In their 2004 paper, Grossman and Ozluk stated that "When working with a spreadsheet, the developer makes a series of choices about what to do and how to do it". Such choices are personal to the developer. This reflects the development approach adopted for the construction of the Sudoku analysis. The layout and methods of analysis of this model are very unusual. This stems from the facts that we are using identities rather than numeric values as the core input to the model and that the model's complexity derives from the overlapping nature of the ranges. The methodology employed is almost equivalent to a mathematician's approach to solving simultaneous equations.

We considered it a worthwhile exercise to compare our approach to the construction of the model and its various elements to academic descriptions of the processes that should be considered. We selected two significant papers [Grossman and Ozluk, 2004] and [Ko et al., 2011]. The later paper considered the approaches to the creation of all types of software by end-users.

The review [Allen, 2013] displayed an unexpectedly good fit between the methods adopted for the construction of the Sudoku model and the academic descriptions. Accordingly, the question arises as to whether the model conforms to expected standards of construction or not. If not, should a new standard be created?

Clearly the structures and methods of construction employed for this model do not conform to the standards that we would apply to the financial and actuarial models. These models form the bedrock of the types of spreadsheets investigated by academia. This lack of conformity is not surprising since the core of spreadsheets under investigation by academics is fundamentally numeric in character rather than spatial.

Spreadsheets that adopt this spatial mode of analysis are rare. There is an example concerning the mapping of the ecology of a national park in the Northern Territories of Australia [McMahon et al., 2010] and other approaches to solving Sudoku puzzles in Excel – for example [Weiss & Rasmussen, 2007]. Also, ACBA (UK) LTD developed a "spatially aware" version of a daybook for its commercial accounting system [ACBA, 2000]. These examples (including the current model) number only four amongst the huge numbers (millions?) of spreadsheets created since their introduction in the 1980s. It is open to argument therefore whether this style of spreadsheet construction and approach deserves its own separate classification.

### 3.2. Complexity in Spreadsheets

In his 2004 paper Bregar discussed potential metrics for evaluating the complexity of spreadsheets. He stated that:

"It has been proven that the complexity of a model or a particular formula represents an important factor to be considered in the process of spreadsheet development, because a complex spreadsheet makes error finding difficult and because errors come in relation with cells that have a high potential for faults."

He examined the impact of IF statements and LOOKUP tables on the reliability of models. He concluded that the conditionality of these constructs, which allowed some paths to be followed and others not, hugely increased the complexity of a spreadsheet from the perspective of a possible metrical analysis.

We examine these concepts in relation to the Sudoku model. Lists of all the formulae employed within the model are available from the ACBA (UK) LTD website [Allen, 2012].

**Relationship between the design requirement and complexity**

Some of the formulae within this model employ complex constructs. The underlying question is whether these constructs are necessary to the model. That in turn depends on both the nature of the model and the nature of the question that the user is seeking to answer.

In this context the model looks relatively simple, but the logical relationships between the internal areas of the model are complex. The model as described here is designed to present this complexity such that a reviewer can evaluate each step in the process.

Since, from a presentational viewpoint, the answer is a representation of the pathway rather than merely the final solution, we needed to consider how much control the design of the system should impose upon the user. This model is designed to demonstrate both the misuse of logic that generates an error and also the correct use of logic. This might be regarded as the human connection.

In order to proceed with the construction of a usable model we sought to limit the potential range of user created errors – see the green coloured input cells in Figures 2, 4 and 5 – which all employ drop down lists.

The majority of the calculations/assessments, subsequent to this user input, remain open to inspection. They are controlled by formulae and/or programmed relationships. This leaves open the question of whether the design of these programmed relationships is accurate. Overall, there is no substantive evidence to suggest we have failed in this programmed approach.

This approach to construction generates a very different perception of the nature and control of a relatively complex model from that proposed by Bregar. The emphasis here is on an open presentation of the complex logic so as to encourage understanding, rather than limiting complexity so as to avoid errors.

**Formula Complexity and Context**

The formula complexity within this environment seeks to ensure that ALL the processes necessary for completing the analysis and creating a new version of the puzzle are indeed complete. In a form-based environment (e.g. a database) this can be processed progressively, whereas within a worksheet one has to create artificial controls. Part of this artifice is to deliberately generate an error (or something that looks like an error).

The following formula is extracted from the bottom right hand cell in Figure 2. It is characterised by the nested IF statements that Bregar calls into question and also includes the artificial error discussed above.

=IF(AND(RowNo<>"",F37=""),"#Error",IF(OR(F37="Pre-Set",F37="Solved",F37="1 of 2"),INDIRECT("L" & (RowNo+7)),IF(F51="Solved",SUM(H40:H48),IF(F51="1 of 2",MID(H49,1,1)&", "&MID(H49,2,1),"."))))

It may be instructive to consider this formula in detail. We need to consider why it has been encapsulated into a single cell and whether we regard this as suitable given the number of nested levels that it incorporates.

**Table 1 – Deconstructing a complex Excel formula**

| Nested IF Statement | Result Type | Commentary |
|---|---|---|
| IF(AND(RowNo<>"",F37=""),"#Error", | Fixed String | Validates that the Sudoku cell under investigation has been identified and prevents any failure to do so going un-noticed |
| IF(OR(F37="Pre-Set",F37="Solved",F37="1 of 2"),INDIRECT("L" & (RowNo+7)), | Numeric Identity | Checks whether the cell has been already either Pre-set or Solved and delivers the identity |
| IF(F51="Solved",SUM(H40:H48), | Number | Adds all the values in range H40:H48 but since only one value has been posted it will be correct |
| IF(F51="1 of 2",MID(H49,1,1)&", "&MID(H49,2,1) | Variable string | Generates a string of which two identities could form the result |
| ,".")))) | Full point string | If none of the previous cases are valid, posts a point string, indicating an unresolved value |

This creation of a deliberate error is designed to prevent such mistakes carrying forward unnoticed. The other four nested elements of formula are also designed to be carried forward into a revised puzzle grid. In these cases they are all potentially valid responses to the analysis. It is very unusual to permit such a mixture of data types within a single cell. In the context discussed here, however, it provides the range of values needed to continue working with the Sudoku puzzle.

## 4. CONCLUSION

Through the medium of the Sudoku model we can visualize approaches to spreadsheet construction which are based on spatial relationships rather than numeric analysis. This model allows us to investigate a range of control measures to manage errors and risks that are not present in conventional number-based models. The model is designed to illuminate errors in logic rather than eliminate them.

We have considered two of the academic methods for describing or categorising spreadsheets (or similar end-user constructs). The methods of construction match, but the end product looks and feels very different to a standard spreadsheet. The question of whether there should be a new category for this style of spreadsheet construct has been posed but left unresolved.

We considered both the use of open spreadsheet structures and the associated complexity in formulae that this appeared to generate. We believe that the structural requirement of any model is very much dependent on its purpose. This in turn influences the complexity of the formulae and relationships employed. Simple constructs are valuable but not a universal panacea.

## BIBLIOGRAPHIC REFERENCES